\documentclass[pra, onecolumn, nofootinbib]{revtex4}
\usepackage{graphicx}
\usepackage{amssymb}
\usepackage{amsmath}
\usepackage{epsfig}
\usepackage{subfigure}
\usepackage{graphics}

\begin{document}
\date{\today}
\title{Wave Refraction in Left-Handed Materials}
\author{Antonia Chimonidou}
\email[email:]{antonia@physics.utexas.edu} \affiliation{The University of Texas at Austin, Center for Complex
Quantum Systems, 1 University Station C1600, Austin TX 78712}
\author{E.~C.~G. Sudarshan}
\affiliation{The University of Texas at Austin, Center for Complex Quantum Systems, 1 University Station C1600,
Austin TX 78712}
\begin{abstract}
We examine the response of a plane wave incident on a flat surface of a medium characterized by simultaneously negative $\epsilon$ and $\mu$ by solving Maxwell's equations explicitly and without making any assumptions on the way. In the literature up to date, it has been assumed that negative refractive materials are necessarily frequency dispersive. We propose an alternative to this assumption by suggesting that the requirement of positive energy density should be relaxed, and discuss the implications of such a proposal. More specifically, we show that once negative energy solutions are accepted, the necessity for frequency dispersion is no longer necessary. Finally, we argue that, for the purposes of discussing negative index materials, the use of group velocity as the physically significant quantity is misleading, and suggest that any discussion involving it should be carefully reconsidered.
\end{abstract}

\maketitle
\section{Introduction}
Refraction is perhaps one of the most fundamental phenomena of optics and electromagnetic wave propagation. When a beam of light is incident on an interface between two different materials, its path is deflected depending on the ratio of the refractive indices $n_1$ and $n_2$ of the two materials. The refractive index of a medium measures how fast it transmits light and how light is bent on entering the material from another medium. The higher the ratio of the indices of refraction of the two media, the slower the propagation and the stronger the deflection. The basis of lenses and imaging lies in the phenomenon of refraction, as any material with an index different than that of its environment will deflect any incoming ray that is not normal to the interface. In addition to the refractive index, the electric permittivity $\epsilon$ and magnetic permeability $\mu$ are commonly used material parameters that describe how materials polarize in the presence of electric and magnetic fields. The electric permittivity determines a material's response to an applied electric field, while the magnetic permeability summarizes how the material reacts to an applied magnetic field.  Maxwell's equations relate these two parameters to the refractive index $n$ as follows:
\begin{equation}
\label{n}
n=\pm\sqrt{\epsilon\mu}
\end{equation}
For materials found in nature, the positive sign of the square root in Eq. (\ref{n}) applies. 

Materials can be classified in terms of the sign of their permittivity $\epsilon$ and permeability $\mu$. Almost all materials encountered in optics, such as glass or water, have positive values for both permittivity $\epsilon$ and permeability $\mu$. However, many metals (such as silver and gold), have negative $\epsilon$ at visible wavelengths. Materials having either (but not both) $\epsilon$ or $\mu$ negative are opaque to electromagnetic radiation (for example surface plasmons), since $n$ becomes imaginary. To have $n$ real, we need to look at materials with the same sign of $\epsilon$ and $\mu$ simultaneously. However, materials with both $\epsilon$ and $\mu$ negative for long wavelengths are not found in nature.

The theoretical implications of materials with simultaneous negative permeability and permittivity were first studied by Victor Veselago, a Russian physicist,  in 1968 \cite{Veselago}. Veselago suggested that the electromagnetic properties of such materials would exhibit interesting phenomena, without violating any fundamental physical laws. Materials like these have since then been given the name ``metamaterials'', or ``left-handed materials'' (LHM for short). Veselago claimed that such materials would be characterized by a negative index of refraction, and would have unique properties such as a reversed Doppler effect, Cherenkov radiation, and even Snell's Law. Since then, a great deal of interest has been paid to these theories \cite{Valanju1, Pendry3, Valanju2, Smith2}.

Even though materials with $\epsilon$ and $\mu$ negative simultaneously are not found in nature, it has been recently suggested that such materials can be fabricated over a finite frequency band by specific electromagnetic design. Such a one-dimensional structure was introduced by Smith \emph{et al.} in \cite{Smith1}. Subsequently, an extension to two dimensions was presented by Shelby  \emph{et al.}, and such an artificial composite was predicted to have an isotropic, negative index of refraction in two dimensions \cite{Shelby1}. Recent technological advances have made this goal realizable. In fact, experimental scattering data at microwave frequencies on a structured metamaterial that exhibits a frequency band where the effective index of refraction is negative, directly confirms Veselago's predictions \cite{Shelby2}. To produce a negative electric permittivity in a particular frequency region, wire elements were used \cite{Pendry1}, while at the same time, non magnetic thin sheets of metal were utilized to create a negative magnetic permeability in an overlapping frequency region \cite{Pendry2}. Many other investigators have since then confirmed negative refraction at microwave frequencies \cite{Houck, Parazzoli}.

In this paper, we study the behavior of a plane electromagnetic wave incident on a metamaterial from vacuum. This problem has been discussed extensively in the literature during the last decade, but we approach the problem in a more fundamental manner, without making any assumptions on the way. We explicitly solve Maxwell's equations and arrive at the physically significant information by matching the boundary conditions at the interface. We specifically concentrate our attention to two of the assumptions put forth in Veselago's original work: the first has to do with the requirement of a positive energy density, which in turn leads to the necessity of frequency dispersion in negative refractive materials, and the second is involved with the use of group velocity as the quantity which describes the behavior of a wave once inside a LHM. In response to these assumptions, we make two proposals: the first is that the positive energy density requirement should be relaxed, and the other is that the use of group velocity as the variable which determines physical information should be abandoned.

This paper is organized as follows: in Sec. II, we solve Maxwell's equations for a plane wave incident on a LHM from vacuum, for both parallel and perpendicular polarizations. Our results are then used to describe the response of the wave once inside the metamaterial. In Sec. III, we suggest why the positive energy density requirement should be relaxed and discuss the implications of such a change. We finish this section with a general discussion on group velocity and its use for the purpose of describing the behavior of a wave in the LHM. Finally, in Sec. IV, we summarize our results and offer some concluding remarks. 

\section{Calculation}
As mentioned above, for naturally occurring materials, the positive sign of the square root in Eq. (\ref{n}) is used. In \cite{Veselago}, Veselago claims that if a medium possesses simultaneously negative $\epsilon$ and $\mu$, this convention must be reversed to satisfy energy considerations: the negative sign of the square root in Eq. (\ref{n}) must be chosen instead. This statement seems to have no direct justification. As an aside, and before we continue to solve Maxwell's equations explicitly, we wish to shed some light to why $n$ becomes negative when both $\epsilon$ and $\mu$ are less than zero, without any energy considerations and by simply considering Snell's law. Snell's law is given by:
\begin{equation}
\label{Snell}
n_1\sin{\theta_1}=n_2\sin{\theta_2},
\end{equation}
where $n_1$ is the index of refraction of the medium that the wave is incident from, say vacuum, and $n_2$ is that of the left-handed medium. As per usual, $\theta_1$ and $\theta_2$ are respectively the angles of incidence and refraction at the interface. Equation (\ref{Snell}) can be written in terms of the wave vector magnitudes as:
\begin{equation}
\label{Snellk}
|\textbf{k}_1|\sin{\theta_1}=|\textbf{k}_2|\sin{\theta_2},
\end{equation}
where we have used the relation:
\[
n=\frac{c|\textbf{k}|}{\omega},
\]
with $k$ being the magnitude of the wave vector describing the wave, and $\omega$ its frequency. 
The sign of the refractive index $n=n_1/n_2$, is determined by the sign of the ratio $|\textbf{k}_1|/|\textbf{k}_2|$, which really depends not on the magnitude, but on the direction of the wave vectors in the two materials. In what follows, we show that the sign of this ratio turns out to be negative for materials described by simultaneously negative $\epsilon$ and $\mu$.

Having made this clarification, we begin our calculation by considering a two-dimensional plane wave and letting the plane of incidence be the x-y plane. The interface between vacuum and the metamaterial is placed at $y=0$. We are interested in the behavior of the wave vector $\textbf{k}$ as the wave enters the left-handed medium from vacuum. For this purpose, we let $\epsilon,\mu>0$ for $y<0$ and $\epsilon,\mu<0$ for $y>0$. There are two possible scenarios: the first is when the electric field is perpendicular to the plane of incidence and the second is when it is parallel to it. In this section, we study these two cases separately and show that the same results are obtained in either case.

\subsection{Electric Field Perpendicular to Plane of Incidence}
We consider here the electric field to be perpendicular to the plane of incidence (the x-y plane). The plane wave can be described in terms of the electric field as:
\begin{eqnarray}
\label{ElectricFieldVector}
\textbf{E}(\textbf{r},t) = 
\begin{cases} 
\textbf{E}_{0I}e^{i(\textbf{k}_I\cdot\textbf{r}-\omega t)}+\textbf{E}_{0R}e^{i(\textbf{k}_R\cdot\textbf{r}-\omega t)} & y<0 \\
\textbf{E}_{0T}e^{i(\textbf{k}_T\cdot\textbf{r}-\omega t)}&   y>0
\end{cases}
\end{eqnarray}
with
\begin{eqnarray}
\nonumber
\textbf{k}_I&=&k_x\hat{\textbf{x}}+k_y\hat{\textbf{y}}\\
\nonumber
\textbf{k}_R&=&k_x\hat{\textbf{x}}-k_y\hat{\textbf{y}}\\
\nonumber
\textbf{k}_T&=&k_x'\hat{\textbf{x}}+k_y'\hat{\textbf{y}}.
\end{eqnarray}
$\textbf{E}_{0I}$, $\textbf{E}_{0R}$ and $\textbf{E}_{0T}$ denote the amplitudes of the incident, reflection and transmission vectors. The wave vectors are related by
\[
k_Iv_1=k_Rv_1=k_Tv_2=\omega,
\]
with $v_1=c$. The fields are perpendicular to the propagation vector $\textbf{k}$. Equation (\ref{ElectricFieldVector}) can be rewritten as:
\begin{eqnarray}
\label{ElectricFieldComponents}
\textbf{E}(\textbf{r},t) =e^{-i\omega t} 
\begin{cases} 
e^{ik_xx}\left[E_{0I}e^{ik_yy}+E_{0R}e^{-ik_yy}\right]\hat{\textbf{z}}& y<0 \\
e^{ik_x'x}\left[E_{0T}e^{ik_y'y}\right]\hat{\textbf{z}}&   y>0
\end{cases}
\end{eqnarray}
To calculate the magnetic field, we employ Faraday's Law, $\nabla\times\textbf{E}=-\frac{\partial}{\partial t}\textbf{B}$:
\begin{equation}
\nonumber
\left.\begin{aligned}
 -\frac{\partial}{\partial t}B_x=\frac{\partial}{\partial y}E_z=(ik_y)e^{-i\omega t}e^{ik_xx}\left[E_{0I}e^{ik_yy}-E_{0R}e^{-ik_yy}\right]&\\
 \frac{\partial}{\partial t}B_y=\frac{\partial}{\partial x}E_z=(ik_x)e^{-i\omega t}e^{ik_xx}\left[E_{0I}e^{ik_yy}+E_{0R}e^{-ik_yy}\right]&
\end{aligned}\right\}
\qquad \text{$y<0$}
\end{equation}
and
\begin{equation}
\nonumber
\left.\begin{aligned}
-\frac{\partial}{\partial t}B_x&=&\frac{\partial}{\partial y}E_z=E_{0T}(ik_y')e^{-i\omega t}e^{ik_x'x}e^{ik_y'y}&\\
\frac{\partial}{\partial t}B_y&=&\frac{\partial}{\partial x}E_z=E_{0T}(ik_x')e^{-i\omega t}e^{ik_x'x}e^{ik_y'y}&
\end{aligned}\right\}
\qquad \text{$y>0$}
\end{equation}
This translates to:
\begin{equation}
\label{BxByBefore}
\left.\begin{aligned}
B_x&=&\left(\frac{1}{\omega}\right)(k_y)e^{-i\omega t}e^{ik_xx}\left[E_{0I}e^{ik_yy}-E_{0R}e^{-ik_yy}\right]&\\
B_y&=&-\left(\frac{1}{\omega}\right)(k_x)e^{-i\omega t}e^{ik_xx}\left[E_{0I}e^{ik_yy}+E_{0R}e^{-ik_yy}\right]  &
\end{aligned}\right\}
\qquad \text{$y<0$}
\end{equation}
and
\begin{equation}
\label{BxByAfter}
\left.\begin{aligned}
B_x&=&\left(\frac{E_{0T}}{\omega}\right)(k_y')e^{-i\omega t}e^{ik_x'x}e^{ik_y'y}&\\
B_y&=&-\left(\frac{E_{0T}}{\omega}\right)(k_x')e^{-i\omega t}e^{ik_x'x}e^{ik_y'y}&
\end{aligned}\right\}
\qquad \text{$y>0$},
\end{equation}
or
\begin{eqnarray}
\label{MagneticFieldComponents}
\textbf{B}(\textbf{r},t) =\frac{1}{\omega}e^{-i\omega t}
\begin{cases} 
e^{ik_xx}\left[(E_{0I}e^{ik_yy}-E_{0R}e^{-ik_yy})(k_y)\hat{\textbf{x}}-(E_{0I}e^{ik_yy}+E_{0R}e^{-ik_yy})(k_x)\hat{\textbf{y}}\right]& y<0 \\
E_{0T}e^{ik_x'x}e^{ik_y'y}[k_y'\hat{\textbf{x}}-k_x'\hat{\textbf{y}}]&   y>0
\end{cases}
\end{eqnarray}
We can use Ampere's Law, $\nabla\times\textbf{H}=\frac{\partial}{\partial t}\textbf{D}$, or $\nabla\times\textbf{B}=\epsilon\mu\frac{\partial}{\partial t}\textbf{E}$, to arrive at the wave equation of the wave. The components are:
\begin{eqnarray}
\epsilon\mu\frac{\partial E_x}{\partial t}&=&-\frac{\partial B_y}{\partial z}\nonumber\\
\epsilon\mu\frac{\partial E_y}{\partial t}&=&\frac{\partial B_x}{\partial z}\nonumber\\
\epsilon\mu\frac{\partial E_z}{\partial t}&=&\frac{\partial B_y}{\partial x}-\frac{\partial B_x}{\partial
y}\nonumber
\end{eqnarray}
The only relevant equation is the third one. Using Eqs. (\ref{ElectricFieldComponents}) and (\ref{MagneticFieldComponents}), we obtain: 
\begin{eqnarray}
\label{Dispersion}
\epsilon\mu\omega^2 = 
\begin{cases} 
k_x^2+k_y^2& y<0 \\
k_x'^2+k_y'^2& y>0
\end{cases}
\end{eqnarray}
Before we can present a more detailed analysis of the behavior of the wave as it gets refracted into the left-handed medium, we calculate the Poynting vector associated with the electromagnetic field of the wave. The Poynting vector can be thought of as the energy flux of the electromagnetic wave in the direction of the energy flow. It is given by $\textbf{S}=\frac{1}{\mu}\textbf{E}\times\textbf{B*}$. For the form of the wave presented above, 
\[
\textbf{S}=\frac{1}{\mu}\left[-E_zB_y^{*}\hat{\textbf{x}}+B_x^{*}E_z\hat{\textbf{y}}\right].
\]
From Eqs. (\ref{ElectricFieldComponents}) and (\ref{MagneticFieldComponents}), the x and y components of the Poynting vector are found to be:
\begin{equation}
\label{SxSyBefore}
\left.\begin{aligned}
S_x&=&-\frac{1}{\mu}E_zB_y^{*}=\frac{1}{\omega}\left(\frac{k_x}{\mu}\right)(E_{0I})^2[1+Re^{2ik_yy}+Re^{-2ik_yy}+R^2]&\\
S_y&=&\frac{1}{\mu}E_zB_x^{*}=\frac{1}{\omega}\left(\frac{k_y}{\mu}\right)(E_{0I})^2[1-Re^{2ik_yy}+Re^{-2ik_yy}-R^2]&
\end{aligned}\right\}
\qquad \text{$y<0$}
\end{equation}
and
\begin{equation}
\label{SxSyAfter}
\left.\begin{aligned}
S_x&=&-\frac{1}{\mu'}E_zB_y^{*}=\frac{1}{\omega}\left(\frac{k'_x}{\mu'}\right)(E_{0T})^2&\\
S_y&=&\frac{1}{\mu'}E_zB_x^{*}=\frac{1}{\omega}\left(\frac{k'_y}{\mu'}\right)(E_{0T})^2&
\end{aligned}\right\}
\qquad \text{$y>0$},
\end{equation}
where $R=E_{0R}/E_{0I}$ and $T=E_{0T}/E_{0I}$.

At this point, it is important to notice that we have neither said anything about the signs of $\epsilon$ or $\mu$, nor have we made any assumptions about the direction of the wave vector inside the left-handed medium. The physically significant information is hidden in the boundary conditions at the interface of the metamaterial. These ask for continuity of the normal components of $\textbf{D}$ and $\textbf{B}$, and similarly for the parallel components of $\textbf{E}$ and $\textbf{H}$. In other words, we ask for continuity in $E_z$, $B_y$ and $H_x$. This gives:
\begin{eqnarray}
(1+R)e^{ik_xx}&=&Te^{ik'_xx}\label{one}\\
k_x(1+R)e^{ik_xx}&=&k'_xTe^{ik_x'x}\label{two}\\
\left(\frac{k_y}{\mu}\right)(1-R)e^{ik_xx}&=&\left(\frac{k'_y}{\mu'}\right)Te^{ik'_xx}\label{three},
\end{eqnarray}
Equations (\ref{one}) and (\ref{two}) taken together lead to the condition:
\begin{equation}
\label{good_condition_y}
k_x=k_x',
\end{equation}
and Eq. (\ref{two}) becomes:
\[
1+R=T.
\]
The reflection and refraction amplitudes are found to be:
\begin{eqnarray}
R&=&\frac{k_y/\mu-k'_y/\mu'}{k'_y/\mu'+k_y/\mu}\label{reflection_1}\\
T&=&\frac{2k_y/\mu}{k'_y/\mu'+k_y/\mu}\label{transmission_2}
\end{eqnarray}
To reveal the behavior of the wave vector as it enters the metamaterial, we make use of the additional requirement that the component of the Poynting vector across the interface is continuous. This last condition can be written as:
\[
\left(\frac{k_y}{\mu}\right)(1-R^2)=\left(\frac{k'_y}{\mu'}\right)T^2,
\]
or
\begin{equation}
\label{good_condition_x}
\frac{1-R^2}{T^2}=\frac{\left(\frac{k'_y}{\mu'}\right)}{\left(\frac{k_y}{\mu}\right)}.
\end{equation}
Here, $\mu=\mu_0>0$ is the permeability of vacuum and $\mu'<0$ is that of the left-handed material. We notice that the left-hand side of Eq. (\ref{good_condition_x}) is positive. As a consequence, since $k_y/\mu>0$ and $\mu'<0$, it is required that $k'_y<0$. Eqs. (\ref{good_condition_y}) and (\ref{good_condition_x}) thus completely determine the direction of the transmitted wave vector $\textbf{k}_T$.  

\subsection{Electric Field Parallel to Plane of Incidence}
For completion, we now turn to the slightly more complicated case in which the electric field is perpendicular to the plane of interface, and show that the results produce conclusions which are identical to those presented in the previous subsection. The setup of the problem remains the same, but we now have to make use of the conditions $\hat{\textbf{E}}_{0I}\cdot\hat{\textbf{k}}_I=\hat{\textbf{E}}_{0R}\cdot\hat{\textbf{k}}_R=\hat{\textbf{E}}_{0T}\cdot\hat{\textbf{k}}_T=0$ in order to calculate the incident, reflection, and transmission vectors. They are given by:
\begin{eqnarray}
\textbf{E}_{0I}&=&E_{0I}(k_y\hat{\textbf{x}}-k_x\hat{\textbf{y}})\nonumber\\
\textbf{E}_{0R}&=&-E_{0R}(k_y\hat{\textbf{x}}+k_x\hat{\textbf{y}})\nonumber\\
\textbf{E}_{0T}&=&E_{0T}(k'_y\hat{\textbf{x}}-k_x'\hat{\textbf{y}})\nonumber
\end{eqnarray}
In terms of the electric field, the wave can be written as:
\begin{eqnarray}
\label{ElectricFieldVector_parallel}
\textbf{E}(\textbf{r},t) = 
\begin{cases} 
E_{0I}(k_y\hat{\textbf{x}}-k_x\hat{\textbf{y}})e^{i(k_xx+k_yy-\omega t)}-E_{0R}(k_y\hat{\textbf{x}}+k_x\hat{\textbf{y}})e^{i(k_xx-k_yy-\omega t)} & y<0 \\
E_{0T}(k'_y\hat{\textbf{x}}-k'_x\hat{\textbf{y}})e^{i(k'_xx+k'_yy-\omega t)}&   y>0
\end{cases}
\end{eqnarray}
or
\begin{eqnarray}
\label{E_parallel_simplified}
\textbf{E}(\textbf{r},t) =e^{-i\omega t} 
\begin{cases} 
e^{ik_xx}\{[E_{0I}e^{ik_yy}-E_{0R}e^{-ik_yy}](k_y)\hat{\textbf{x}}-[E_{0I}e^{ik_yy}+E_{0R}e^{-ik_yy}](k_x)\hat{\textbf{y}}\}  & y<0 \\
e^{ik'_xx}E_{0T}(k'_y\hat{\textbf{x}}-k'_x\hat{\textbf{y}})e^{ik'_yy}&   y>0
\end{cases}
\end{eqnarray}
The magnetic field is found to be:
\begin{eqnarray}
\label{MagneticFieldComponents_parallel}
\textbf{B}(\textbf{r},t) =\frac{e^{-i\omega t}}{-\omega}
\begin{cases} 
e^{ik_xx}(k^2_x+k^2_y)(E_{0I}e^{ik_yy}+E_{0R}e^{-ik_yy})\hat{\textbf{z}}& y<0 \\
e^{ik'_xx}E_{0T}(k'^2_x+k'^2_y)e^{ik'_yy}\hat{\textbf{z}}&   y>0
\end{cases}
\end{eqnarray}
The wave equation can be calculated for this case and, as expected, it is found to be identical to Eq. (\ref{Dispersion}). The components of the Poynting vector are given by: 
\begin{equation}
\label{SxSyBefore_new}
\left.\begin{aligned}
S_x&=&\frac{1}{\mu}E_yB_z^{*}=\frac{1}{\omega}\left(\frac{k_x}{\mu}\right)(k^2_x+k^2_y)(E_{0I})^2[1+Re^{2ik_yy}+Re^{-2ik_yy}+R^2]&\\
S_y&=&-\frac{1}{\mu}E_xB_z^{*}=\frac{1}{\omega}\left(\frac{k_y}{\mu}\right)(k^2_x+k^2_y)(E_{0I})^2[1+Re^{2ik_yy}-Re^{-2ik_yy}-R^2]&
\end{aligned}\right\}
\qquad \text{$y<0$}
\end{equation}
and
\begin{equation}
\label{SxSyAfter_new}
\left.\begin{aligned}
S_x&=&\frac{1}{\mu'}E_yB_z^{*}=\frac{1}{\omega}\left(\frac{k'_x}{\mu'}\right)(k'^2_x+k'^2_y)(E_{0T})^2&\\
S_y&=&-\frac{1}{\mu'}E_xB_z^{*}=\frac{1}{\omega}\left(\frac{k'_y}{\mu'}\right)(k'^2_x+k'^2_y)(E_{0T})^2&
\end{aligned}\right\}
\qquad \text{$y>0$}
\end{equation}
The boundary conditions ask for continuity in $E_x$, $D_y$, and $H_z$. These give:
\begin{eqnarray}
k_y(1-R)e^{ik_xx}&=&k'_yTe^{ik'_xx}\label{one_2}\\
\epsilon k_x(1+R)e^{ik_xx}&=&\epsilon' T(k'_x)e^{ik'_xx}\label{two_2}\\
\frac{1}{\mu}(k^2_x+k^2_y)(1+R)e^{ik_xx}&=&\frac{1}{\mu'}(k'^2_x+k'^2_y)Te^{ik'_xx}\label{three_2},
\end{eqnarray}
Equations (\ref{two_2}) and (\ref{three_2}) taken together state that:
\begin{equation}
\label{k_y_2}
\frac{k_x}{k'_x}=\frac{\epsilon'\mu'}{\epsilon\mu}\frac{(k^2_x+k^2_y)}{(k'^2_x+k'^2_y)},
\end{equation}
and this completely determines the sign of $k'_x$: since $\epsilon\mu,\epsilon'\mu'>0$, $k'_x$ must be positive. Similarly, the continuity of the component of the Poynting vector across the interface can be used to determine the sign of $k'_y$. It reads:
\begin{equation}
\label{k_x_2}
\frac{k_y}{\mu}\frac{(k^2_x+k^2_y)}{(k'^2_x+k'^2_y)}\frac{(1-R^2)}{T^2}=\frac{k'_y}{\mu'}
\end{equation}
Once again, $k'_y<0$. 

\begin{figure}
\setlength{\unitlength}{0.4mm}
\begin{center}
\begin{picture}(100,250)(-50,-130)
\put(145,-15){\small $x$}
\put(-15,125){\small $y$}
\put(100,-15){\small $\epsilon,\mu>0$}
\put(100,15){\small $\epsilon',\mu'<0$}
\put(0,0){\vector(1,0){150}}
\put(0,0){\line(-1,0){150}}
\put(0,0){\vector(0,1){130}}
\put(0,0){\line(0,-1){130}}
\put(-100,-100){\vector(1,1){50}}
\put(-50,-50){\line(1,1){50}}
\put(-150,-100){\vector(1,1){50}}
\put(-100,-50){\line(1,1){50}}
\put(-70,-50){\small $\textbf{k}_I$}
\put(-120,-50){\small $\textbf{S}_I$}
\put(-100,100){\vector(1,-1){50}}
\put(-50,50){\line(1,-1){50}}
\put(-150,100){\line(1,-1){50}}
\put(-50,0){\vector(-1,1){50}}
\put(-70,50){\small $\textbf{k}_T$}
\put(-120,50){\small $\textbf{S}_T$}
\put(0,0){\vector(1,-1){50}}
\put(50,-50){\line(1,-1){50}}
\put(-50,0){\vector(1,-1){40}}
\put(-10,-40){\line(1,-1){60}}
\put(10,-50){\small $\textbf{S}_R$}
\put(60,-50){\small $\textbf{k}_R$}
\end{picture}
\end{center}
\caption{Ray diagram for the behavior of the wave as it propagates into a left-handed medium placed at $y=0$. The ``ray" is shown in terms of the Poynting vector as well as the wave vector. \label{RayDiagram}}
\end{figure}
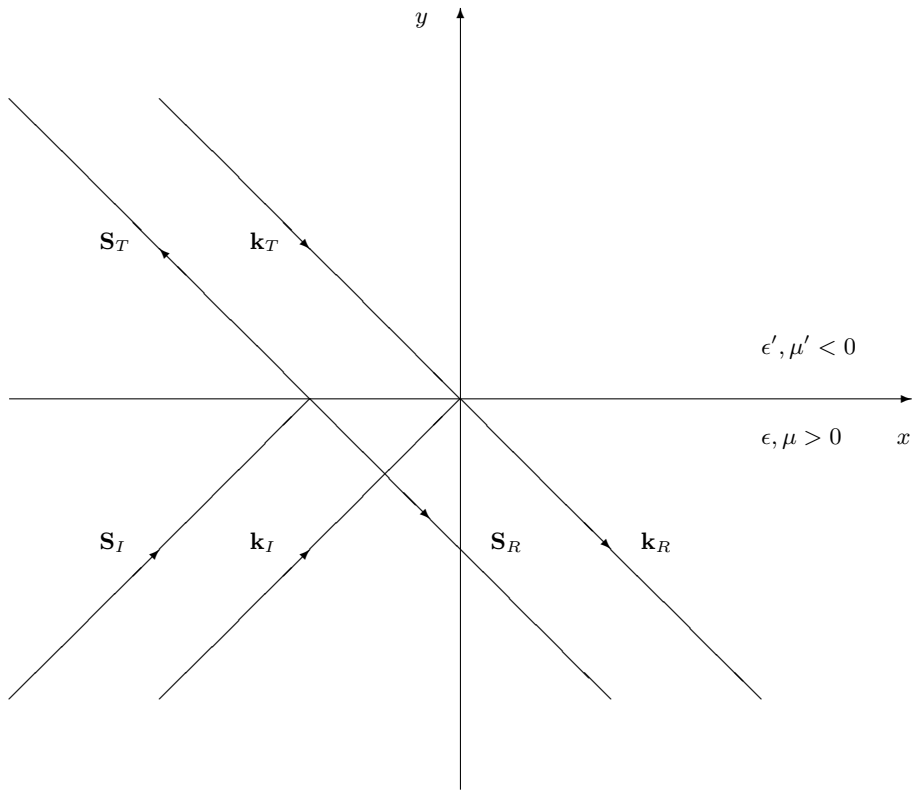

The reflection and transmission amplitudes are found to be:
\begin{eqnarray}
R&=&\frac{\epsilon' k'_xk_y-\epsilon k_xk'_y}{\epsilon k_xk'_y+\epsilon' k'_xk_y}\nonumber\\
T&=&\frac{2\epsilon k_xk_y}{\epsilon k_xk'_y+\epsilon'k'_xk_y}\nonumber
\end{eqnarray} 
A great deal of physically important information is contained in the ratios $\frac{k_x'}{\mu'}$, and $\frac{k_y}{\mu'}$ appearing in Eqs. (\ref{SxSyAfter}) and (\ref{SxSyAfter_new}). Since $\mu$ is positive, Eqs. (\ref{SxSyBefore}) and (\ref{SxSyBefore_new}) state that the Poynting vector is parallel to the wave vector in the region $y<0$. However, as the sign of the permeability changes from positive to negative as the wave moves from vacuum to the left-handed material, the Poynting vector becomes antiparallel to the wave vector in the region $y>0$, as shown by Eqs. (\ref{SxSyAfter}) and (\ref{SxSyAfter_new}). This result is not new. In fact, Veselago arrived at the same conclusion by using different arguments. The situation is depicted in Fig. (\ref{RayDiagram}). Notice that, as expected, the energy flux is away from the interface for $y>0$.

\section{Analysis}
In all literature available up to date, it has been assumed that negative refractive materials are necessarily frequency dependent. In this section, we propose an alternative to this assumption. Consider the energy density of the electromagnetic wave in a non-dispersive medium (Eq. (22) in \cite{Veselago}):
\begin{equation}
\label{energy_density_nodispersion}
W=\epsilon E^2+\mu H^2
\end{equation}
In his original paper, Veselago poses an argument that has to do with the sign of Eq. (\ref{energy_density_nodispersion}). Specifically, he requires that $W$ be positive, and continues to suggest that, since simultaneously negative values of $\epsilon$ and $\mu$ would make $W<0$, Eq. (\ref{energy_density_nodispersion}) should be replaced with the expression for the energy density in a dispersive medium (Eq. (23) in \cite{Veselago}): 
\begin{equation}
\label{energy_density_dispersion}
W=\frac{\partial(\epsilon\omega)}{\partial\omega}E^2+\frac{\partial(\mu\omega)}{\partial\omega}H^2.
\end{equation}
According to Veselago, in order to have $W>0$ in Eq. (\ref{energy_density_dispersion}), it is required that $\frac{\partial(\epsilon\omega)}{\partial\omega}>0$, and $\frac{\partial(\mu\omega)}{\partial\omega}>0$. More importantly, because of the positive energy density requirement, negative refraction can be realized only when there is frequency dispersion. As a final remark, Veselago claims that the partials in Eq. (\ref{energy_density_dispersion}) do not in general mean that $\epsilon$ and $\mu$ cannot be simultaneously negative, but for them to hold, it is necessary that $\epsilon$ and $\mu$ are frequency dependent.

There are many situations in physics where energy assumes negative values. Two of the most well-known examples are those arising from the Klein-Gordon and the Dirac relativistic wave equations. Both these equations admit negative energy solutions which are not realized in nature. This gives an apparently fatal difficulty, since an external perturbation can cause a particle with energy greater than its rest mass energy $mc^2$ to make a radiative transition to a state of negative energy less than $mc^2$. In 1930, Dirac proposed a way out of this difficulty by suggesting that, for electrons, there are states in nature that admit negative energy, but they are all full. Since, according to the Pauli exclusion principle, only one electron is allowed in each state,
transitions from positive to negative energy states are only allowed if there is a mechanism that empties the already full negative
states. In this picture, the vacuum state consists of an infinite sea of electrons filling all the negative energy states and all the
positive energy states are empty. It is possible for an electron in the negative sea to acquire energy greater than $2mc^2$ and make a transition to a positive energy state with some value greater than $mc^2$. This transition leaves behind a \emph{hole} in the sea of negative-energy states. Measured with respect to the vacuum, this hole appears to have a positive charge $(+e)$ and positive energy; it can therefore be interpreted as a \emph{positron}. This is how Dirac predicted the existence of positrons, which were indeed discovered in 1933 by C.D.Anderson.

The Dirac field with the negative energy sea all filled, has difficulties with relativistic invariance. The problem presents itself when the negative energy states are reinterpreted as holes. For an extensive discussion see \cite{Friedrichs}. However, there are other examples in which negative energy presents itself. Let's consider the case in which there is a region in space in which the energy density $W$ associated with the motion of particles is negative. Such a question was considered in \cite{Sudarshan} for the case of tachyons, particles capable of overcoming the light barrier. Consider being in a frame $S$ which is at rest. Also suppose that in that frame, a tachyon is observed to be emitted by a source and absorbed a while later by a sink. It was shown that, for an observer in some other frame $S'$ moving with velocity $w$ with respect to frame $S$, these particles may appear to have negative energy. In addition, to the observer in the $S'$ system, the negative-energy particle will appear to have been absorbed first and emitted later. At that time, this phenomenon raised serious objections to the possible existence of tachyons. To resolve this difficulty, the authors proposed the \emph{reinterpretation principle}. According to this principle, a negative-energy particle that has been absorbed first and emitted later, is nothing else but a positive-energy particle emitted first and absorbed later, a perfectly normal situation. 

It also turns out that, for the case of tachyons, the velocity is antiparallel to the momentum. But this is precisely what we observe in negative refractive index materials: the Poynting vector is antiparallel to the wave vector. Quite similarly to the case of tachyons, it looks like, inside the LHM, the wave hits the interface before it is emitted by its source. However, by the reinterpretation technique, we can think of a negative-energy wave propagating in one direction as a positive-energy wave propagating in the opposite direction. 

Allowing the existence of negative energy solutions has consequences which contradict Veselago's predictions. Namely, allowing $W$ to be negative means that the partials in Eq. (\ref{energy_density_dispersion}) are no longer restricted to be positive. More importantly, however, Eq. (\ref{energy_density_nodispersion}) need not be replaced by Eq. (\ref{energy_density_dispersion}), and the requirement that the medium be dispersive if negative refraction can occur is no longer valid. In fact, it makes sense that this is so. After all, what happens in certain regions of a material in which the absorption bands are far enough such that the region in-between them is essentially non-dispersive for a quie-monichromatic pulse? Once negative energy solutions are accepted, it seems like there is no fundamental reason justifying the rejection of negative refraction in non-dispersive materials. 

\subsection{Dispersion, and the Vagueness of Group Velocity}
We continue our discussion with a review of the fundamentals of the phenomenon of dispersion and discuss the vagueness of the use of group velocity to describe the behavior of a wave inside a metamaterial. In Sec. II of \cite{Veselago}, Veselago states that in negative index materials, $\textbf{S}$ and $\textbf{k}$ are antiparallel, and continues to argue that such materials must have negative group velocity. The assumption here is that the so-called group velocity $\textbf{v}_g$ is in the same direction as the Poynting vector, which is in turn antiparallel to $\textbf{k}$ and also the phase velocity $\textbf{v}_p$. He continues to say that $\textbf{v}_p$ is opposite to the energy flux, which he takes to be $\textbf{S}$ or $\textbf{v}_g$ interchangeably. In the discussion that follows, we claim that the use of group velocity as presented by Veselago is in general quite an unclear one, and that the notion of group velocity should be used very carefully, if at all. 

When a ray traverses a prism, the change of the incident angle $\theta$ for the various wavelengths can be measured with a spectrometer. The rate of change $d\theta/d\lambda$ is the angular dispersion of the prism. It can be conveniently represented as the product of two factors, by writing:
\begin{equation}
\label{angular_dispersion}
\frac{d\theta}{d\lambda}=\frac{d\theta}{dn}\frac{dn}{d\lambda}
\end{equation}
The first factor in Eq. (\ref{angular_dispersion}) can be evaluated by geometrical considerations alone, while the second factor is a characteristic property of the prism material. This is what we refer to as the material's \emph{dispersion}. We consider here some of the known facts about the variation of $n$ with $\lambda$. It is found that if one were to plot a graph of $n$ versus $\lambda$ for a range of different optical materials, that curve would differ in detail from one material to the next, but all curves would have the same general shape. Curves representative of the so-called \emph{normal dispersion}, are characterized by the following facts: $n$ increases with increasing $\lambda$, the rate of increase $dn/d\lambda$ rises at shorter wavelengths, the larger the index of refraction of a certain substance, the steeper $dn/d\lambda$ is for a given wavelength, and the curve of one substance cannot in general be obtained from that of another substance by a simple change in the scale of the ordinates. Considering the last statement, or the fact that there is no simple relation between the different curves, the dispersion of different materials is said to be \emph{irrational}. The magnitude of $n$ is in general quite different for various substances, but its change with wavelength shows the characteristics described above.

If the index of refraction is measured for substances like quartz in the infrared region of the spectrum,  the dispersion curve begins to deviate significantly from that which describes the so-called normal dispersion. More specifically, the measured value of $n$ decreases more and more rapidly with increasing $\lambda$, until it reaches a region in the infrared where light ceases to be transmitted at all. This region of selective absorption is known as the absorption band, and its position is characteristic of the material in question. Since the substance does not transmit radiation of the wavelength in the absorption band, $n$ cannot usually be measured in this region. The index of refraction is found to be higher for the long wavelength side of the absorption band, and as one moves further away from the band, $n$ decreases rapidly at first and more slowly later. This phenomenon is known as \emph{anomalous dispersion}. 

Anomalous dispersion was first observed with materials whose absorption bands fall in the visible region of the spectrum. However, it was soon discovered that transparent substances like glass and quartz possess regions of selective absorption in the infrared and ultraviolet regions, and therefore show anomalous dispersion in those areas. It turns out that no substance exists which does not exhibit anomalous dispersion at some wavelengths, making the term ``anomalous'' inappropriate. This phenomenon, far from being anomalous, is perfectly general. The so-called ``normal'' dispersion is seen only in the region between two absorption bands, and fairly far removed from them. Nonetheless, the term ``anomalous dispersion'' has been retained for historical reasons. 

It is not uncommon for some positive index materials to exhibit regions in which $\textbf{v}_p>0$ and $\textbf{v}_g<0$ (references). In this sense, there is nothing special about negative index materials, contrary to Veselago's assertion that left-handed materials are materials with negative group velocity. To make things more transparent, we recall that the phase and group velocity of a wave are related by:
\begin{equation}
\label{pos_or_neg}
v_g=v_p+k\frac{dv_p}{dk}=v_p-\lambda\frac{dv_p}{d\lambda},
\end{equation}
where the relation $k=2\pi/\lambda$ has been used to arrive to the second part of the equation above. Equation (\ref{pos_or_neg}) can also be expressed in terms of the refractive index $n$ as:
\begin{equation}
\label{v_g_n}
v_g=\frac{\partial\omega}{\partial k}=\frac{c}{{\rm{Re}}(n)+\omega\frac{\partial{\rm{Re}}(n)}{\partial\omega}}.
\end{equation}
In the fictitious case of the absence of dispersion, the group velocity is identical to the phase velocity in both magnitude and direction. However, the sign of group velocity can be positive or negative depending on the sign of the second term in the denominator of Eq. (\ref{v_g_n}) and its magnitude with respect to ${\rm{Re}}(n)$. It should thus be expected that the relative sign between $\textbf{v}_p$ and $\textbf{v}_g$ vary in different regions of the same material, irrespective of the sign of its refractive index. In fact, all four sign combinations between $\textbf{v}_p$ and $\textbf{v}_g$ have been verified experimentally for a negative index metamaterial \cite{Dolling}. It is important to note at this point that nowhere in the above discussion has the Poynting vector been mentioned. To conclude our claim that the use of group velocity is inappropriate, we further note the connection between the Poynting vector and the group velocity:
\begin{equation}
\label{S_and_v_g}
\textbf{S}=W\cdot\textbf{v}_g.
\end{equation}  
Clearly, the relative sign between $\textbf{S}$ and $\textbf{v}_g$ depends on the sign of $W$. The calculations of Sec. II show that $\textbf{S}>0$ at all times, signifying that the energy flux is away from the interface inside the metamaterial. In other words, the Poynting vector is positive - otherwise, no signal would be transmitted through the sample. The fact that $\textbf{v}_g$ can be either positive or negative in a dispersive medium verifies the suggestion that negative energy solutions should not be rejected. Even though there exists no material in which dispersion is absent entirely, there surely must be regions for which dispersion is so small that it can be thought of as negligible. According to Veselago, negative refraction is not allowed in those regions. Based on fundamental physical reasoning, we claim that this is simply not true.

Since no assumptions of the nature of the medium were made in deriving Eq. (\ref{Dispersion}), we are in a position to study the behavior of the wave vector as a function of frequency in the region of interest for any medium of our choice. Below, we present two cases: in the first, we consider the fictitious case of a non-dispersive medium in which $\epsilon$ and $\mu$ are frequency independent, and in the second,  we consider the case of a medium with a simple frequency dependence of the form $\epsilon(\omega)=\epsilon_0(1-\frac{\omega^2_p}{\omega^2})$ and $\mu(\omega)=\mu_0(1-\frac{\omega^2_p}{\omega^2})$, where $\omega_p$ is the constant plasma frequency. We focus our attention to the region $y>0$ for the case where the electric field in perpendicular to the plane of interface, such that $k_x=k'_x$.
 
\begin{figure}[h]
\begin{center}
\resizebox{13cm}{8cm}{\includegraphics{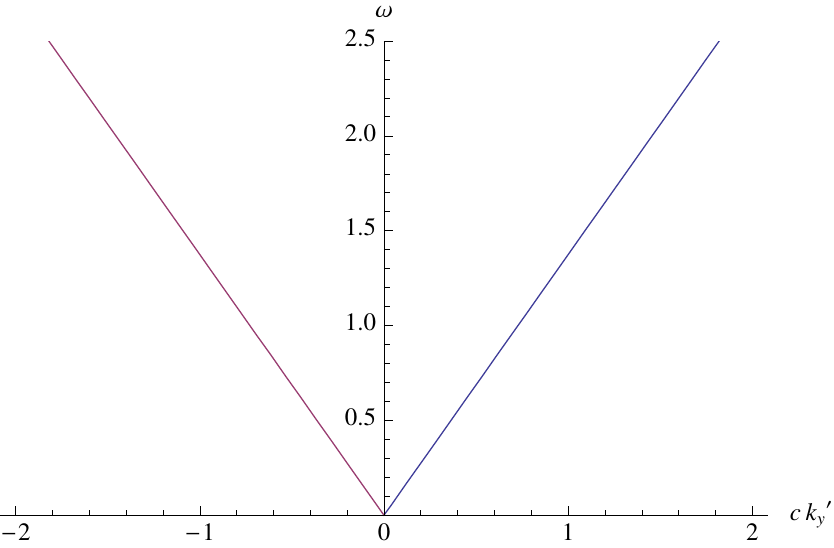}}
\end{center}
\caption{Behavior of the y-component of the refracted part of the wave vector as a function of frequency for $x>0$ and $\epsilon,\mu$ negative constants less than 1. The wave is incident at an angle of $30^{\circ}$.}\label{fig:NoDispersion}
\end{figure}

In the case of non-dispersive media, equation (\ref{Dispersion}) becomes:
\begin{equation}
\label{NoDispersion}
k'^2_yc^2=\omega^2\left[\frac{\epsilon\mu}{\epsilon_0\mu_0}-\sin^2{\theta_i}\right],
\end{equation}
with $\epsilon$, $\mu$ negative constants and less than 1. Here, $\theta_i$ is the incident angle of the wave, and $k'_x$ has been replaced with $\frac{\omega}{c}\sin{\theta_i}$. Equation (\ref{NoDispersion}) has  two possible solutions:
\begin{equation}
k'_yc=\pm\omega\sqrt{\frac{\epsilon\mu}{\epsilon_0\mu_0}-\sin^2{\theta_i}}\label{NonDispersiveSolutions}
\end{equation}
The behavior of the y-component of the wave vector as a function of frequency is a simple linear relation and is shown in Fig. \ref{fig:NoDispersion} for an incident angle of $\theta_i=30^{\circ}$ and positive $\omega$.

Following the results of the calculations of Sec. II, the physical region of interest is that in which $k'_y<0$. As an aside, note that for this region, $\textbf{v}_p$ and $\textbf{v}_g$ are negative and parallel. This is of no surprise, of-course, since in the absence of dispersion, the group velocity of the wave is precisely its phase velocity. At the same time, the energy flux is positive. Equation (\ref{S_and_v_g})  requires that $W$ be negative for non-dispersive media.

\begin{figure}[h]
\begin{center}
\resizebox{13cm}{8cm}{\includegraphics{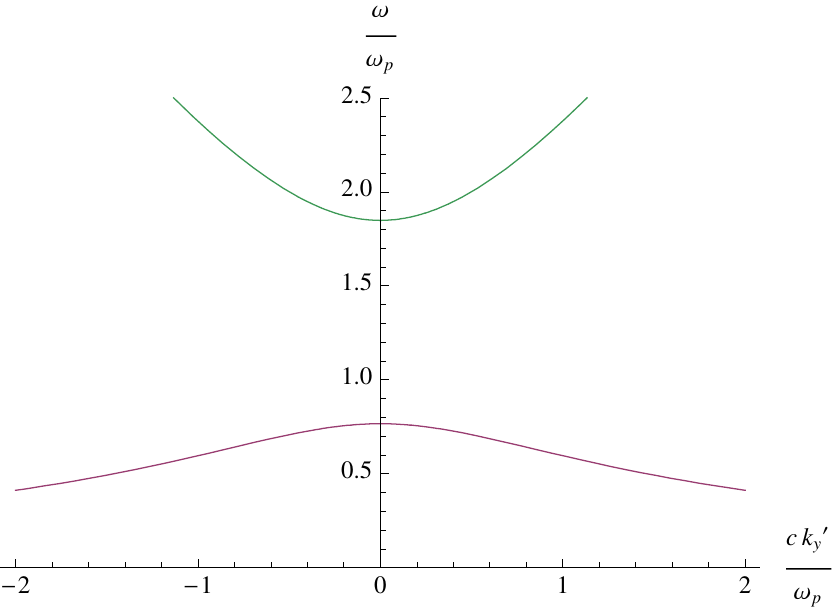}}
\end{center}
\caption{Behavior of the y-component of the refracted part of the wave vector as a function of frequency for $x>0$ and $\frac{\epsilon}{\epsilon_0}=\frac{\mu}{\mu_0}=1-\frac{\omega_p^2}{\omega^2}$. The wave is incident at an angle of $45^{\circ}$.} \label{fig:2DDispersion}
\end{figure}

We proceed to study the same problem in the case where $\epsilon$ and $\mu$ are described by a simple frequency dependence of the form $\epsilon(\omega)=\epsilon_0(1-\frac{\omega^2_p}{\omega^2})$ and $\mu(\omega)=\mu_0(1-\frac{\omega^2_p}{\omega^2})$, where $\omega_p$ is the constant plasma frequency. Under this assumption, equation (\ref{Dispersion}) takes the form:
\begin{equation}
\label{DispersiveEquation}
k'^2_yc^2=\omega^2\left[\left(1-\frac{\omega^2_p}{\omega^2}\right)^2-\sin^2{\theta_i}\right]
\end{equation}
Once again, there are two possible solutions to Eq. (\ref{DispersiveEquation}):
\begin{equation}
k'_yc=\pm\omega\sqrt{\left(1-\frac{\omega^2_p}{\omega^2}\right)^2-\sin^2{\theta_i}}\label{DispersiveSolutions}
\end{equation}
A plot of the real values of the y-component of the wave vector as a function of frequency for the case where $\epsilon$ and $\mu$ have this particular frequency dependence is shown in Fig. \ref{fig:2DDispersion}.

Fig. \ref{fig:2DDispersion} displays all four solutions of Eq. (\ref{DispersiveSolutions}). The upper branch has $\frac{\omega}{\omega_p}>1$ and corresponds to $\epsilon$ and $\mu$ positive. Similarly, the lower branch has $\frac{\omega}{\omega_p}<1$ and corresponds to $\epsilon$ and $\mu$ negative. Again, we are interested in the region in which $k'_y<0$. Figure \ref{fig:2DDispersion} shows that for positive index materials $\textbf{v}_p$ and $\textbf{v}_g$ are negative and parallel, while for negative index materials, $\textbf{v}_p<0$ and $\textbf{v}_g>0$. Again, the energy flux is positive in both cases. 

\section{Concluding Remarks}

We have examined the response of a monochromatic plane electromagnetic wave incident on a negative index material from vacuum by explicitly solving Maxwell's equations and without making any assumptions on the nature of the medium. The physically significant information was explored by matching boundary condition at the interface. We suggested that negative energy solutions should not be rejected on the grounds of being unphysical, and presented the implications arising from such a change. We specifically discussed the fact that by accepting negative energy solutions, we are allowing negative refraction to occur in regions within a material possessing negligible dispersion. Further, we showed that all possible sign combinations between phase and group velocity of a wave should be expected in all materials, regardless of the sign of the refractive index. For completion, we solved the wave equation for two extreme cases, one in which dispersion was absent, and another in which the high-frequency limit was assumed.    

The original prescription for a sub-wavelength array of thin metallic wires combined with resonant metallic rings has been extensively investigated, and negative refraction at microwave frequencies has been confirmed by several investigators. In addition, a great deal of effort has been put in trying to create a material with a negative refractive index for visible light. In fact, in 2000, Notomi showed that negative refraction should be expected to occur in photonic crystals for a certain range of frequencies \cite{Notomi1}. Photonic crystals are periodic structures built on the scale of the optical wavelength, which allow only certain wavelengths to pass through. Essentially, they are structures in which the refractive index can be controlled by their band structure. The nice thing about photonic crystals, other than the fact that they are built within the visible region, is that they have significantly reduced losses. Many researchers have already explored negative refraction and Cherenkov radiation using photonic crystals \cite{Notomi2, Luo1, Luo2, Foteinopoulou, Parimi1, Parimi2, Cubukcu1, Cubukcu2}. Just like for the case of negative index metamaterials, all possible sign combinations between phase and group velocity were observed in photonic crystals \cite{Mojahedi1}. One problem with the original metamaterial structures presented is that they were not suitable for practical microwave applications because of their excessive loss and narrow bandwidth. Caloz \emph{et al.} introduced a transmission line approach of left-handed materials and proposed an artificial left-handed transmission line with microstrip components including interdigital capacitors and shorted stub inductors \cite{Caloz}. Since then, researchers have arrived at the circuit equivalent of a negative index material using transmission lines \cite{Liu, Eleftheriades, Grbic, Omar}. It is very important to note here that experiments involving transmission-line devices demonstrate only moderate dispersion with no noticeable increase in absorption. It is understandable that materials in which dispersion is negligible would be impossible to study experimentally. However, these additional approaches to negative refraction involving photonic crystals and circuit models indicate a promising future as far as the arguments on the necessity of dispersion put forth in this paper.


\begin{thebibliography}{99}


\bibitem{Veselago} V.~G.~Veselago, {Sov.\ Phys.\ Usp.} {\bf 10}, 509 (1968)

\bibitem{Valanju1} P.~M.~Valanju, P.~M.~Walser, and A.~P.~Valanju, {Phys.\ Rev.\ Lett.} {\bf 88}, 187401 (2002)

\bibitem{Pendry3} J.~B.~Pendry and D.~R.~Smith, {Phys.\ Rev.\ Lett.} {\bf 90}, 029703 (2003)

\bibitem{Valanju2} P.~M.~Valanju, P.~M.~Walser, and A.~P.~Valanju, {Phys.\ Rev.\ Lett.} {\bf 90}, 029704 (2002)

\bibitem{Smith2} D.~R.~Smith, D.~Schurig, and J.~B.~Pendry, {Appl.\ Phys.\ Lett.} {\bf 81}, 2713 (2002)

\bibitem{Smith1} D.~R.~Smith, W.~J.~Padilla, D.~C.~Vier, S.~C.~Nemat-Nasser, and S.~Schultz, {Phys.\ Rev.\ Lett.} {\bf 84},
4184 (2000)

\bibitem{Shelby1} R.~A.~Shelby, D.~R.~Smith, S.~C.~Nemat-Nasser, and S.~Schultz, {Appl.\ Phys.\ Lett.} {\bf 78}, 489 (2001)

\bibitem{Shelby2} R.~A.~Shelby, D.~R.~Smith, and S.~Schultz, {Science} {\bf 292}, 77 (2001)

\bibitem{Pendry1} J.~B.~Pendry, A.~J.~Holden, W.~J.~Steward, and I.~Youngs, {Phys.\ Rev.\ Lett.} {\bf 76}, 4773 (1996)

\bibitem{Pendry2} J.~B.~Pendry, A.~J.~Holden, D.~J.~Robbins, and W.~J.~Steward, {IEEE\ Trans.\ Microwave\ Theory\ Tech.} {\bf 47}, 2075 (1999)

\bibitem{Houck} A.~A.~Houck, J.~B.~Brock, and I.~Chuang, {Phys.\ Rev.\ Lett.} {\bf 90}, 137401 (2003)

\bibitem{Parazzoli} C.~G.~Parazzoli, R.~B.~Greegor, K.~Li, B.~E.~C.~Koltenbah, and M.~Tanielian, {Phys.\ Rev.\ Lett.} {\bf 90}, 107401 (2003)

\bibitem{Dolling} G.~Dolling, C.~Enkrich, M.~Wegener, C.~Soukoulis, and S.~Linden, {Science} {\bf 312}, 892 (2006)

\bibitem{Friedrichs} K.~O.~Van Friedrichs in {\it Mathematical aspects of the quantum theory of fields}, (New York, Interscience Publishers, 1953)

\bibitem{Sudarshan}   O.~M~Bilaniuk, and E.~C.~G.~Sudarshan, {Physics Today} {\bf 22}, 43 (1969) 

\bibitem{Notomi1} M.~Notomi, {Phys.\ Rev.\ B} {\bf 62}, 10696 (2000)

\bibitem{Notomi2} M.~Notomi, {Optical \ and \ Quantum \ Electronics} {\bf 34}, 133 (2002)

\bibitem{Luo1} C.~Luo, S.~G.~Johnson, J.~D.~Joannopoulos, and J.~B.~Pendry {Phys.\ Rev.\ B} {\bf 65}, 201104(R) (2002)

\bibitem{Luo2} C.~Luo, M.~Ibanescu, S.~G.~Johnson, and J.~D.~Joannopoulos, {Science} {\bf 299}, 368 (2003)

\bibitem{Foteinopoulou} S.~Foteinopoulou, E.~N.~Economou, and C.~M.~Soukoulis, {Phys.\ Rev.\ Lett.} {\bf 90}, 107402 (2003)

\bibitem{Cubukcu1} E.~Cubukcu, K.~Aydin, E.~Ozbay, S.~Foteinopoulou, and C.~M.~Soukoulis, {Nature} {\bf 423}, 604 (2003)

\bibitem{Cubukcu2} E.~Cubukcu, K.~Aydin, E.~Ozbay, S.~Foteinopoulou, and C.~M.~Soukoulis, {Phys.\ Rev.\ Lett.} {\bf 91}, 207401 (2003)

\bibitem{Parimi1} P.~V.~Parimi, W.~T.~Lu, P.~Vodo, and S.~Sridhar, {Nature} {\bf 426}, 404 (2003)

\bibitem{Parimi2} P.~V.~Parimi, W.~T.~Lu, P.~Vodo, J.~Sokoloff, J.~S.~Derov, and S.~Sridhar, {Phys.\ Rev.\ Lett.} {\bf 92}, 127401 (2004)

\bibitem{Mojahedi1} M.~Mojahedi, K.~J.~Malloy, G.~V.~Eleftheriades, J.~Woodley, and Y.~Chiao, {IEEE\ Journal\ of\ Selected\ Topics \ in \ Quantum \ Electronics} {\bf 9}, 30 (2003)

\bibitem{Caloz} C.~Caloz, H.~Obake, T.~Iwai, and T.~Itoh, {USNC/URSI \ National\ Radio \ Science \ Meeting, \ San \ Antonio, \ TX} (2002)

\bibitem{Liu} L.~Liu, C.~Caloz, C.~Chang, and T.~Itoh, {J.\ Appl.\ Phys.} {\bf 92}, 5560 (2002)

\bibitem{Eleftheriades} G.~V.~Eleftheriades, A.~K.~Iyer, and P.~C.~Kremer, {IEEE\ Trans.\ Microwave\ Theory\ Tech.} {\bf 50}, 2702 (2002)

\bibitem{Grbic} A.~Grbic, and G.~V.~Eleftheriades, {Phys.\ Rev.\ Lett.} {\bf 92}, 117403 (2004)

\bibitem{Omar} O.~F.~Siddiqui, S.~J.~Erickson, G.~V.~Eleftheriades, and M.~Mojahedi, {IEEE\ Trans.\ Microwave\ Theory\ Tech.} {\bf 52}, 1449 (2004)




\end{thebibliography}
\end{document}